\documentstyle[12pt]{article}
\newcommand{\beq}{\begin{equation}\label}
\newcommand{\eeq}{\end{equation}}
\newcommand{\p}{\partial}

\newcommand{\f}{\frac}
\newfont{\testb}{msbm10}
\begin{document} 

\title{Zamolodchikov's C-theorem and phase transitions\thanks{The work was presented as a poster at Summer School ``Confinment, Duality and Non-perturbative aspects of QCD'', June 24 - July 4, 1997, Cambridge, UK.}} 
\author{Maxim Zabzine\thanks{e-mail: zabzin@vanosf.physto.se}\\
\begin{small}\textit{Institute of Theoretical Physics, University of Stockholm}\end{small}\\
\begin{small}\textit{Box 6730, S-113 85  Stockholm, Sweden}\end{small}\\
\begin{small}\textit{and}\end{small}\\
\begin{small}\textit{Department of Theoretical Physics, St Petersburg University}\end{small}\\ 
\begin{small}\textit{Ulyanovskaia 1, St Petersburg, 198904, Russia}\end{small}}
\maketitle
.\vskip -12.0cm
\hfill USITP-97-11
\vskip 12.0cm
.\vskip -12.5cm
\hfill hep-th/9707064
\vskip 12.5cm

\begin{bf}Abstract \end{bf}: We discuss the possibility of generalizing some aspects of the C-theorem within two different approaches, the conventional RG and the Wilson RG flows. We show that the original Zamolodchikov's theorem is related to the  existence of the phase transitions in finite temperature QFT. We present some arguments related to the holomorphic property of the low energy Wilson effective action.\\

\newpage  

\section{Introduction}\label{1}

A important tool for modern quantum field theory (QFT) and statistical mechanics is the idea of the renormalization group (RG) and effective dynamics. The RG describes the fundamental property of nature, the decoupling of physical phenomena at different energy scales. Originally renormalization was just a means for removing infinities in perturbative calculations in QFT. Later Wilson took the idea of physical cut-off at a very large scale seriously and defined the exact RG flow \cite{Wilson}. These two difinitions of RG flow give us two ways to introduce a momentum scale. The first leads to the Gell-Mann-Low beta function and Callan-Symanzik equation, the second to the exact RG flow on the space of lagrangians. There is a relation between these two RGs, but it is a non-trivial task to make explicit in all generality.

Intuitivelly one thinks of the RG flow as describing a ``irreversability''. More precisely, when we  pass from a smaller scale to a large scale, we average over some irrelevant degrees of freedom (they become integration variables). In this situation we expect the existence of entropy-like function along RG flow. 

Zamolodchikov established the existence of a dimensionless function with monotonic property  along RG trajectories, usually called the C-function \cite{Zam}. He proved this within the framework of the conventional RG, with the beta functions satisfying the Callan-Symanzik equations. The C-function is a function of the coupling constants associated with marginal and relevant operators. Attempts to generalize Zamolodchikov's argument to dimensions large than two in straightforward ways fail essentially because the rotation group is no longer abelian. 

The intuition for the Zamolodchikov's C-theorem is ussually expressed in terms of `thinning out of the degrees of freedom'. But the speculation about dergees of freedom is more appropriate within the framework of Wilson's exact RG, where we study the flows of all couplings associated with marginal, relevant and irrelevant operators. Recently Periwal shown that the free energy decreases along Wilson RG trajectories, in a dimension independed way \cite{Per}. The decreasing of the free energy along Wilson RG flow does not imply the C-theorem on submanifold of marginal and relevant couplings. It seems to be that an entropy-like function can be introduced only along Wilson RG flow. This function will be a function of all coupling constants including irrelevant directions. The restriction of this function to a submanifold of relevant and marginal couplings is a very non-trivial task. 

Here we discuss the possibilities of a generalization of some aspects of the C-theorem within two different approaches, the conventional RG and the Wilson RG flows.  

The content of the letter is as follows: In section \ref{2} we discuss the case of two dimensional QFT. Zamolodchikov's theorem can be generalized in a straightforward manner to finite temperature 2D QFT. This generalization gives us a possibility to study the thermodynamical aspects of the C-theorem. In section \ref{3} we explain why two dimensional case is special, in the sense that C-function can be universal only in this case. Using thermodynamical arguments, we show that Zamolodchikov's function for QFT in an arbitrary number of dimensions  has non universal properties and that monotonicity of this function is destroyed by phase transitions. In section \ref{4} we discuss some evidence for the existence of an entropy-like function along the the RG flow and consider Periwal's result for the exact Wilson RG flow. The  behavior of the free energy is related to  the behavior the Wilson effective action along the RG flow. 
   
We briefly consider the possible relation between analyticity of the free energy at low temperature and analyticity of the Wilson effective action at low energy.                        

\section{The C-theorem for 2D QFT}\label{2}

Zamolodchikov proved a theorem for two dimensional QFT which says that there exists a function of coupling constants $C(g_{1}, g_{2},..., g_{n})$, which is monotonically decreasing along the RG flow
\beq{a1}
\Lambda \frac{dC}{d\Lambda} = - \beta_{i}(g)\frac{\p}{\p g_{i}}C(g) \leq 0
\eeq
and stationary for conformally invariant theories,
\beq{a2}
\beta_{i}(g) = 0 \,\,\,\,\Longleftrightarrow \,\,\,\, \f{\p C}{\p g_{i}} = 0,
\eeq
where it takes the value of the Virasoro central charge \cite{Zam}. Zamolodchikov's proof is very simple  and uses the conditions of renormalizability, positivity, the translational and rotational symmetries, and certain special properties of a 2D conformal field theory.

We would like to consider a finite temperature variant of the C-theorem. To find such a theorem it is best to start from conformal field theory. The finite temperature correction to the free energy density $\cal{F}$ in confromal field theory is given by 
\beq{a3}
\begin{cal}
F
\end{cal}(T) = \begin{cal} F \end{cal}(0) - \f{\pi c_{0}}{6} T^{2}
\eeq
where $c_{0}$ is the central charge of corresponding Virasoro algebra and $T$ is the temperature. Let us generalize this expression to off-critical QFT with some  ultraviolet (UV) cut-off $A$,
\beq{a4}
\begin{cal} F \end{cal} (\lambda, T, A) = \begin{cal} F \end{cal} (\lambda, 0, A) - \f{\pi}{6} T^{2} c(\lambda, T, A),
\eeq
where $\lambda$ is a set of coupling constants. The free energy density does not acquire an anomalous dimension under renormalization, so $c(\lambda, T, A)$ needs no substractions. This implies
\beq{a5}
c(\lambda, T, A) = C(g, \f{\Lambda}{T})
\eeq
where $g$ is a set of renormalizable copuling constants and $\Lambda$ is a renormalization scale. In (\ref{a5}) $C(g, \Lambda/T)$ is finite when the cut-off is removed keeping $g$ fixed. The dependence on $\Lambda$ and $T$ has to be as shown in (\ref{a5}) since $c$ is dimensionless. The function $C$ satisfies the Callan-Symanzik equation
\beq{a6}
(\Lambda \f{\p}{\p \Lambda} + \beta_{i}(g) \f{\p}{\p g_{i}}) C(g, \f{\Lambda}{T}) = 0
\eeq
with Gell-Mann-Low functions
\beq{a7}
\beta_{i}(g) = \Lambda \f{\p g_{i}}{\p \Lambda} 
\eeq
These equations can be solved as usual to yield $C = C(g(\Lambda/T))$. For further detailes of the different  proofs of the finite temperature C-theorem we refer the reader to \cite{zab}. So within the framework of finite temperature 2D QFT we can introduce a positive function of the renormalizable coupling constants $C(g_{1}, g_{2},..., g_{n})$ which is non-increasing along a RG trajectories and non-deacreasing along temperature trajectory
\beq{a8}
\Lambda \f{dC}{d\Lambda} = -T\f{dC}{dT} = -\beta_{i}(g)\f{\p C}{\p g_{i}} \leq 0
\eeq
and which is stationary only at fixed points. At the critical fixed points, the 2D QFT becomes conformally invariant and the value of $C$ at these points is the same as the corresponding central charge.

We have to make some comments about the finite temperature version of the C-theorem. The existence of the C-function in finite temperature QFT is equivalent to the existence of the C-function in standard zero temperature QFT. This becomes clear by considering different proofs of the finite temperature C-theorem \cite{zab}. Every proof in zero temperature QFT can be straightforwardly generalized to finite temperature QFT and vice versa. The C-function defined above (\ref{a4}) can be represented as two point function of stress tensor \cite{mm}.    

\section{Thermodynamical properties of C-function}\label{3}

We have proved that within the finite temperature 2D QFT we can introduce C-function. Let us re-express the finite temperature C-theorem in terms of the thermodynamical functions. The C-function is defined as in (\ref{a4}). The positivity of the C-function is equivalent to the positivity of entropy
\beq{b1}
C(T) \geq 0 \,\,\,\,\Longleftrightarrow \,\,\,\, -\f{\p \begin{cal} F \end{cal}}{\p T} = \begin{cal} S \end{cal} \geq 0
\eeq
where $\cal{S}$ is the entropy density. Positivity of the entropy is postulated in thermodynamics, but can be derived in statistical mechanics. The monotonic property of the C-function can be expressed in terms of free energy density
\beq{b2}
T\f{dC}{dT} \geq 0 \,\,\,\, \Longleftrightarrow \f{\p \begin{cal} F \end{cal}}{\p T} \leq \f{2}{T} (\begin{cal} F \end{cal} (T) - \begin{cal} F \end{cal} (0))
\eeq
where we use Nernst's theorem $\begin{cal} S \end{cal} (0) = 0$. If we represent the free energy density in the standard way in terms of the internal energy density $\cal E$ and the entropy density $\cal S$
\beq{b3}
\begin{cal} F \end{cal} (T) = \begin{cal} E \end{cal} (T) - T\begin{cal} S \end{cal} (T)
\eeq
then the condition (\ref{b2}) can be rewritten as
\beq{b4}
\begin{cal} S \end{cal}(T) \leq \f{2}{T} (\begin{cal} E \end{cal} (T) - \begin{cal} E \end{cal} (0)).
\eeq
This inequality for entropy density is not a consequence of the thermodynamical laws. Thermodynamics requires that the entropy is bounded from below but not from above. The inequality is true only for finite temperature 2D QFT and is related to the fact of the triviality of one dimensional statistical mechanics. There are no phase transitions at nonzero temperature, i.e. no long range order for $T>0$. From statistical mechanics we know that a point of phase transition is defined as point where the free energy is not-analytic\footnote{The free energy describing the pure phase is analitical function of the temperature. In point of phase transition this property is broken. This statement have been proved for large class of lattice systems. Now, one understand the analyticity of the free energy at low temperature as a paradigm in mathematical statistical mechanics \cite{f}, \cite{h}. To generalize this result for QFT is not-trivial task, because we do not know the  properties of the limiting procedures (from lattice to field system).}. So in finite temperature 2D QFT the free energy is an analytical function of the temperature and can be represented as a Taylor series. We see that this fact directly is related to the existence of the C-function. 

We can generalize the definition of the C-function (\ref{a4}) to finite temperature QFT in arbitrary number $D$ of dimensions
\beq{b5}
C(T) = \f{6}{\pi T^{D}}[\begin{cal} F \end{cal}(0) - \begin{cal} F \end{cal} (T)]
\eeq
The function C is dimensionless and positive. The monotonic property of C implies the condition
\beq{b6}
\begin{cal} S \end{cal}(T) \leq \f{D}{(D-1)T}(\begin{cal} E \end{cal} (T) - \begin{cal} E \end{cal} (0))
\eeq
As we have discussed this kind of conditions is not a consequence of the thermodynamical laws. We can not define a monotonic function along a temperature trajectory as a dimensionless part of finite temperature correction to the free energy density. In a general situation the condition (\ref{b6}) can be realised only at low temperature where it is possible that the internal energy density $\cal E$ dominates $T\cal{S}$ in the free energy density. So we can introduce the C-function  for an arbitrary QFT and expect a ``low temperature'' (near IR point) generalization of Zamolodchikov's C-theorem
\beq{b7}
\Lambda \f{dC}{d\Lambda} = -T \f{dC}{dT} \leq 0,\,\,\, T<T_{c},\,\,\, \Lambda > \Lambda_{c}
\eeq
and phase transitions destroy the monotonic behavior of the C-function. 

We have started from original Zamolodchikov's the C-theorem and have considered the natural generalization within framework of the finite temperature two dimensional QFT. The Zamolodchikov's C-function and finite temperature C-function have the same nature. They are related to two point function of stress tensor. Using finite temperature analog of the C-theorem, we have shown that Zamolodchikov's theorem implies the specific enequality for entropy, which can be satisfied only in finite temperature 2D QFT (one dimensional statistical system). We see that original formulation of Zamolodchikov's theorem is related to the special properties of two dimensions such as abelian rotation group and no phase transition (analyticity of the free energy). 

If we want to satisfy our intuition we have to try to find the entropy-like function along RG flow in another place, not two point function of stress tensor.  

\section{Wilson RG flow and C-theorem}\label{4}

Let us show some evidence for existence of the monotonic function along RG flow. At the begining we are not going to discuss the nature of RG flow (conventional or WIlson). Consider the formal manipulations with the free energy
\beq{c1}
F(g, T, A) = - \log \int D\phi\,\,\, e^{S[\phi]}
\eeq
which is well defined object at some UV cut-off $A$. $T$ is temperature and $g$ is a set of coupling constants. Consider the following infinitestimal transformation 
\beq{c2}
x_{\mu} \rightarrow x_{\mu} + \xi_{\mu} x_{\mu},\,\,\,\,\,\,\,|\xi_{\mu}| \ll 1
\eeq
where $x_{\mu}$ denotes the space-time coordinates on $\begin{bf} R \end{bf}^{D-1} \times [0, 1/T]$ and no summation over repeated indexes. The change in the free energy can be represented in terms of the stress tensor $\begin{cal} T \end{cal}_{\mu \nu}$
\beq{c3}
\delta F = - <\delta S> = \int\limits_{0}^{1/T} dx_{0} \int\limits_{\begin{bf} R \end{bf}^{D-1}} d^{D-1}x <\begin{cal} T \end{cal}_{\mu \nu}> \p_{\mu}\xi_{\nu}
\eeq
where we mean the summation over repeated indexes. On the other hand the change in $F$ is due to its scale dependence on the inverse temperature $1/T$ and the UV cut-off (length). So we can read off
\beq{c4}
T\f{\p F}{\p T} = - \int\limits_{0}^{1/T} dx_{0} \int\limits_{\begin{bf} R \end{bf}^{D-1}} d^{D-1}x <\begin{cal} T \end{cal}_{00}> = - \int\limits_{0}^{1/T} dx_{0} <H>
\eeq
where $H$ is the Hamiltonian of the system. In an unitary theory we have the monotonic property of the free energy along temperature flow.

We know that at the end of all calculations the free energy have to depend only from mass scale $\Lambda$ of theory  and temperature $T$. The mass scale appears only effectively, because RG phylosophy requares that total derivative over mass scale from free energy equales zero. This kind of equations are called the RG equations. In standard RG, the renormelization point plays the role of mass scale. In the Wilson RG, the cut-off plays this role. Usualy one separate the full  dependence from mass scale to explicit  and  dependence through coupling constants.  

From our definition (\ref{c1}) of the free energy  we can have only dependence from dimensionless relation $\Lambda/T$. This fact implies 
\beq{c5}
\Lambda \f{\p F}{\p \Lambda} = - T \f{\p F}{\p T} \geq 0
\eeq
We see that the free energy is monotonic along RG flow. The function $(F(0)-F(\Lambda/T))$ is monotonically decreasing along the RG flow. The phase transition does not destroy this property. But questions about how to introduce the mass scale in the theory and which RG flow we mean is still open.

We know two main prescriprion to introduce mass scale. The first is conventional RG prescription and by effective action we understand one-particle-irreducible vertices. The second approach is the Wilson exact RG with another understanding of effective dynamics, where one studies the flows of all couplings in a cut-off theory. The cut-off is present at all stages since  one has irrelevant operators of arbitrary high dimension in effective action. This cut-off plays the role of mass scale in our theory. Resently Periwal proved the monotonic property of the free energy along Wilson RG trajectories. Perival's result\footnote{The Periwal's result has a different sign compared to (\ref{c5}). This is because we define the free energy differently. Periwal's definition is $F = - \log \int D\phi e^{-S[\phi]}$. This difference does not change the physics. The motivation for the definition (\ref{c1}) is to have agreement with previous work \cite{zab}.} agrees with (\ref{c5})
\beq{c6}
<\Lambda \f{\p}{\p \Lambda} S_{int}> = \sum \Lambda \f{\p}{\p \Lambda} g^{i}(\Lambda) <\int \Phi_{i}> \equiv \sum \beta^{i}_{W} \p_{i}F > 0
\eeq
where the free energy consider in zero temperature QFT and free energy is function of $\Lambda/\Lambda_{0}$. $\Lambda_{0}$ is scale where the theory is defined by some concrete lagrangian. $\{\Phi_{i}\}$ is a complete set of scalar operators, including the identity. The manipulation in expression (\ref{c6}) is well defined, by the definition of the Wilson RG , the free energy is left invariant under changes of $\Lambda$, provided the couplings in $S_{int} \equiv \sum g^{i}(\Lambda) \int \Phi_{i}$ are appropriately adjusted. So the changes of the free energy are just changes of the couplings which have to be of the opposite sign as the change due to the explicit presence of the cut-off. In this sense the free energy is defined as function of coupling constants corresponding the irrelevant, relevant and marginal operators. The important remark is that equality (\ref{c6}) have sense perturbatively, as well as non-perturbatively. We have very nice interpritation that the behavior of free energy is related to the behavior of the Wilson effective action along RG flow. 

The manipulation like  (\ref{c6}) is ill defined within framework the conventional RG prescription. From perturbative point of view, we can think about $F(\Lambda/T)$ as monotonic function along RG flow. But we can not say anything about non-perturbative part of flow and the behavior of the free energy have not any physical interpretation like we have in case of the Wilson RG flow.      

Let us make a few remarks about Periwal's result. The irrelevant coupling has dimension in mass units $d<0$, so the coupling at cut-off $\Lambda$ is related to the coupling at $\Lambda_{0}$ by $g'=g(\Lambda_{0}/\Lambda)^{d}$. The intuition tell us that near IR point the Wilson RG flow collapses expotentially down onto the space of renormalazable couplings corresponding relevant and marginal operators. On this submanifold, one expects that the Wilson RG and standard RG flow are closely related. So the irrelevant part in (\ref{c6}) can be small enough such as free energy will be the monotonic function at submanifold of renormalizable couplings
\beq{c7}
\sum \beta^{i}_{W,ir} \p_{i}F + \sum \beta^{i}_{W,rel,mar} \p_{i}F >0\,\,\, \Longrightarrow \,\,\,\sum \beta_{rel, mar}^{i} \p_{i} F>0
\eeq
We see that this fact indicates the possibility of the low energy (near IR point) generalization of original C-theorem. This agrees with thermodynamical consideration.

From statistical mechanics we know that the free energy at low temperature describes the pure phase and is holomorphic function of the temperature. This fact implies the holomorphic property for the free energy as a function of mass scale $\Lambda$ and then we have
\beq{c8}
\Lambda \f{\p}{\p \Lambda} F = - <\Lambda \f{\p}{\p \Lambda} S_{eff}>
\eeq
where $S_{eff}(\Lambda)$ is an effective action in Wilson sense. We see some indication of holomorphic property of the Wilson effective action at low energy. This was rigorously proved for the supersymmetric case \cite{shifman}. Now we have a hope that it is true in general.

\section{Conclusions}\label{5}

We have started from original formulation of the Zamolodchikov's C-theorem in two dimensional QFT. In straightforward manner we have generalized the C-theorem to finite temperature two dimensional QFT. Using this result, we have shown that the C-theorem implies special inequality for entropy which can be satisfied only in two dimensional QFT. In arbitrary number of dimensions this inequality can be satisfied only at low temperature (low energy) and phase transition destroy this inequality. Our result is that two point function of stress tensor can not have universal properties in arbitrary number of dimensions.

We have presented some simple argument that free energy is monotonic along RG flow. But we have remarked that question how to introduce the mass scale is still open. Resently Perival proved that free energy have monotonic property along Wilson RG flow. The behavior of free energy have nice interpritation in terms of the Wilson effective action. It seems to be that the language of the Wilson RG is more natural for thermodynamical speculations.

There are a lot of questions left. We do not discuss the problem of gradient RG flow \cite{mp}. There are still unclear relations between Wilson and standard RGs, so it is difficult to discuss connection between the possible generalization of the C-theorem within framework of these two differnt approaches.   

\section{Acknowledgements}

I am deeply grateful to Ulf Lindstr\"om for many useful discusions and comments. The work  was supported by the grant of the Royal Swedish Academy of Sciences. I would like to thank NATO ASI for the possibility to attend the Summer School ``Confinment, Duality and Non-perturbative aspects of QCD'', June 23 - July 4, 1997, Cambridge, UK.\\


\begin{thebibliography}{99}

\newcommand{\np}{Nucl.\ Phys.\ }
\newcommand{\pr}{Phys.\ Rev.\ }
\newcommand{\cmp}{Commun.\ Math.\ Phys.\ }
\newcommand{\pl}{Phys.\ Lett.\ }

\bibitem{Wilson} K.G.Wilson, Phys.Rev. B4 (1971) 3174, 3184;\\
K.G.Wilson and J.G.Kogut, Phys.Reports 12 (1974) 75.
\bibitem{Zam} A.B. Zamolodchikov, JETP Lett. {\bf 43} (1986) 731;
Sov. J. Nucl. Phys. {\bf 46} (1987) 1090.
\bibitem{Per} V.Periwal, Mod.Phys.Lett., A10 (1995) 1543.
\bibitem{zab} M.Zabzine, preprint USITP-97-05, hep-th/9705015.
\bibitem{mm} N.E.Mavromatos and J.L.Miramontes, Phys.Lett. B226 (1989) 291.
\bibitem{f} J.Fr\"ohlich, Bulletin of the American Mathematical Society, V81(2) (1978) 165.
\bibitem{h} Phase transition and critical phenomena, edited by C.Domb and M.S.Green (Academic Press, London, New York, 1972).
\bibitem{shifman} M.A.Shifman and A.I.Vainshtein, Nucl.Phys. B277 (1986) 456;\\
M.A.Shifman and A.I.Vainshtein, Nucl.Phys. B359 (1991) 571.
\bibitem{mp} R.C.Myers and V.Perival, preprint PUPT-1567, hep-th/9611132.
\end{thebibliography}
\end{document}